\documentclass[aps,prb,groupedaddress,twocolumn]{revtex4}

\usepackage{graphicx}
\usepackage{amsmath}
\usepackage{amsfonts}
\usepackage{amssymb}

\usepackage{subfigure}

\usepackage{color}

\begin{document}

\title{Polarization dependence of phonon influences \\ in exciton-biexciton quantum dot systems}

\author{M. Gl\"assl$^1$} \email[]{martin.glaessl@uni-bayreuth.de}
\author{V.~M.~Axt$^{1}$}

\affiliation{$^{1}$Institut f\"{u}r Theoretische Physik III,
Universit\"{a}t Bayreuth, 95440 Bayreuth, Germany}

\date{\today}

\begin{abstract}

We report on a strong dependence of the phonon-induced damping of Rabi dynamics
in an optically driven exciton-biexciton quantum dot system on the polarization
of the exciting pulse. While for a fixed pulse intensity the damping is maximal
for linearly polarized excitation, it decreases with increasing ellipticity of
the polarization. This finding is most remarkable considering that the carrier-phonon
coupling is spin-independent.
In addition to simulations based on a numerically exact real-time path integral approach,
we present an analysis within a weak coupling theory that allows for analytical
expressions for the pertinent damping rates. 
We demonstrate that an efficient coupling to the biexciton state is of central
importance for the reported polarization dependencies. 
Further, we discuss influences of various
system parameters and show that for finite biexciton binding energies 
Rabi scenarios differ qualitatively from the widely studied two-level dynamics.

\end{abstract}

\maketitle

\section{INTRODUCTION}\label{intro}

Investigations devoted to the carrier dynamics of optically driven quantum dots
(QDs) with or without an additional coupling to a cavity constitute a highly
active field of physics. \cite{ulrich:11,roy:11,vagov:07,ramsay:10,ramsay:10b,
mccutcheon:11,ulhaq:12,kaer:10,wu:11,simon:11,glaessl:12b}
In the limit of strong electronic confinement and for excitations with arbitrarily
polarized laser pulses, the electronic structure of a QD can be modeled as
a four-level system, consisting of the ground state (no electron-hole pair),
two energetically almost degenerate single exciton states with orthogonal
spin polarizations and the biexciton state.
Such exciton-biexciton systems are attractive for many innovative 
technologies, including the realization of sources for on-demand entangled
or squeezed photons \cite{benson:00,gywat:02,hohenester:02,dousse:10} or applications
in the field of quantum information theory. \cite{troiani:00,biolatti:00,chen:01} 
The major obstacle for coherent manipulations is decoherence, and much effort
has been devoted to characterize or mimimize decoherence processes in optically
driven QDs. \cite{krummheuer:02,borri:02,vagov:04,foerstner:03,machnikowski:04,
kruegel:05,axt:05,vagov:07,hodgson:08,ramsay:10,ramsay:10b,mccutcheon:11,
glaessl:12b, lueker:12} Recent experimental
studies of excitonic Rabi rotations \cite{ramsay:10,ramsay:10b} identified the pure
dephasing coupling of carrier states to bulk like acoustic phonons as the principal
source of dephasing in self-assembled
QDs and confirmed theoretically predicted results for phonon-coupled two-level systems,
such as a renormalization of the Rabi frequency. \cite{foerstner:03,kruegel:05} 
Moreover, first evidence for the expected undamping of Rabi
rotations at high pulse areas \cite{vagov:07,glaessl:11} has been shown
by observing a roll-off behavior in the oscillation amplitude. \cite{ramsay:10b} 

Obviously, the dissipative dynamics of a four-level system accounting not only for
single exciton but also for biexciton excitations can develop much more manifold 
than that of a two-level system and is in view of possible applications of special interest.
Most of the existing literature on the phonon impact on the carrier dynamics
of optically driven QDs, however,
deals with electronic two-level systems and comparatively little is known
about the more complex exciton-biexciton dynamics. In Refs.~\onlinecite{axt:05b}
and \onlinecite{kruegel:07} analytic
expressions have been derived for the limit of ultrashort excitations. 
Two-photon Rabi oscillations between the ground and the biexciton state
have been studied experimentally in Ref.~\onlinecite{stufler:06} and
the phonon influence on this scenario has been analyzed theoretically
within a perturbative approach in Ref.~\onlinecite{machnikowski:08}.
Very recently, we have demonstrated that a numerically exact
real-time path integral approach previously used for two-level systems
\cite{vagov:07, glaessl:11, glaessl:11b} can be also implemented for four-level
systems paving the way towards investigations in the regimes of high temperatures,
strong carrier-phonon couplings and long times for arbitrary optical drivings.
\cite{glaessl:12} In that work dark superpositions were identified as a general 
mechanism that strongly modifies the quantum dissipative relaxation,
in particular at long times, and various related dynamical features have been
discussed, including a crossover between two qualitatively different relaxation
scenarios. \cite{glaessl:12}

Here, we analyze in detail the impact of acoustic phonons on  
the dynamics on shorter time-scales than in Ref.~\onlinecite{glaessl:12}
concentrating on Rabi scenarios and signals most convenient to be studied in experiments.
In particular, we study the dynamics for different intensities and 
polarizations of the driving laser field.
As the exciton-phonon coupling is independent of the spin, it might be suspected,
that the phonon influence does not depend on the polarization of the optical
excitation which creates well defined superpositions of carrier spins.
Nevertheless, we shall demonstrate that in contrast to this expectation, 
the phonon-induced damping does strongly depend on the polarization of
the exciting pulse. Keeping the pulse intensity fixed, the damping is most
pronounced for linear polarization, while for elliptical polarizations it
becomes the more reduced the closer the polarization is to the circular limit. 

The paper is organized as follows. In Sec.~\ref{theory} we introduce the model,
we present a short summary of the path integral approach used and comment on
the challenges that are faced when accounting for four electronic levels within
this numerically exact formalism. In addition, we introduce a weak coupling 
theory that allows for approximate but explicit expressions for the pertinent
damping rates. This perturbative approach is described in more detail
in the appendix and most clearly reveals 
the origin of the polarization dependence of the phonon-induced damping.
The latter is presented in Sec.~\ref{results}, where we also study in detail
the dynamics for linearly polarized excitations and discuss the influence of 
pulse intensity, temperature and magnitude of the biexciton binding energy.
Special emphasis is put on comparisons with results for two-level systems.
Finally, we summarize our results in Sec.~\ref{conclusions}.

\section{THEORY}\label{theory} 

\subsection{Model}\label{model}

We consider a strongly confined GaAs QD with spin degenerate electronic particle
states coupled to a continuum of acoustic phonons and driven by laser light. The
corresponding Hamiltonian can be written as
\begin{equation}
H = H_{\rm{dot}} + H_{\rm{ph}} + H_{\rm{dot-ph}} + H_{\rm{dot-light}} \, ,
\label{eq_Hfull}
\end{equation}
where $H_{\rm{dot}}$ describes the electronic structure of the QD, $H_{\rm{ph}}$
is the free phonon Hamiltonian, $H_{\rm{dot-ph}}$ represents the carrier-phonon
and $H_{\rm{dot-light}}$ the carrier-light coupling. As in the strong confinement
limit the electronic single particle states are energetically well separated,
we can safely neglect a Coulomb-induced mixing of states with different single
particle energies and concentrate on the topmost valence and lowest conduction
band states. Neglecting excitonic states that are not coupled to other electronic
states via the carrier-light interaction, our system comprises four states.
Besides the ground state $|\rm{G}\rangle$ without electron-hole pairs we account for
the two single exciton states $|+\rangle$ and $|-\rangle$ with a total angular
momentum of $\pm 1$, and the biexciton state $|\rm{B}\rangle$. Then, $H_{\rm{dot}}$ reads
\begin{align}
H_{\rm{dot}} &= \hbar \, \Omega  \left( | + \rangle \langle + | + | - \rangle \langle - | \right)
             + \left( 2\hbar \Omega - \Delta \right) | \rm{B} \rangle \langle \rm{B} | \\ \notag 
             & \quad + V_{\rm{ex}} \left( | + \rangle \langle - | + | - \rangle \langle + | \right) \, .
\label{eq_Hdot}
\end{align}
We take the unexcited ground state as the zero of energy, $\hbar\Omega$ defines
the ground state exciton transition energy, $\Delta$ denotes the biexciton binding
energy resulting from electrostatic Coulomb interactions and $V_{\rm{ex}}$ the
electron-hole exchange interaction. For simplicity, we consider spherical dots with
envelope functions given by the ground state solution of a harmonic potential, i.~e.
\begin{align}
  \psi_{e(h)}(r) = \frac{1}{\pi^{3/4} a_{e(h)}^{3/2}} \, \exp\left(-\frac{r^2}{2a_{e(h)}^{2}}\right) \, ,
\end{align}
where $a_{e(h)}$ denote the localization lengths of electrons and holes, respectively.

To model the carrier-phonon interaction we concentrate on pure dephasing couplings
\cite{mahan:90, krummheuer:02}
of the electronic states to bulk like longitudinal acoustic (LA) phonons via the 
deformation potential that for strongly confined GaAs QDs has been identified as
the dominant dephasing mechanism \cite{vagov:04,ramsay:10}. 
With $b^{\dagger}_{\bf q}$ denoting the creation operator of a LA phonon with wave
vector ${\bf q}$ and energy $\hbar \omega _{\bf q}$, $H_{\rm{ph}}$ and
$H_{\rm{dot-ph}}$ are then given by
\begin{equation}
H_{\rm{ph}} = \hbar\sum_{\bf{q}}\omega_{\bf q}b_{\bf q}^{\dagger}b_{\bf q} \, ,
\label{eq_Hphonon}
\end{equation}
\begin{equation}
H_{\rm{dot-ph}} = \sum_{\bf{q}} \sum_{\nu} \left( \gamma_{\bf{q}}^{\ast} b_{\bf{q}} + \gamma_{\bf{q}} b_{\bf{q}}^{\dagger}
                                              \right) n_{\nu} |\nu\rangle \langle \nu | \, .
\label{eq_Hdotphonon}
\end{equation}
The dispersion is assumed to be linear, $\omega_{\bf{q}} = v_c |{\bf{q}}|$, where
$v_c=5110\,\rm{m/s}$ is the sound velocity, $n_{\nu}$ is the number of electron-hole pairs present
in the state $|\nu\rangle$ (i.~e., $n_{\rm{G}} \!=\! 0$, $n_+ \!=\! n_- \!=\! 1$ and
$n_{\rm{B}} \!=\! 2$) and $\gamma_{\bf{q}}$ denote the exciton-phonon coupling constants.
We stress, that the carrier-phonon coupling as given in Eq.~(\ref{eq_Hdotphonon})
is independent of the carrier spin. Besides, it should be noted that the factorization
in Eq.~(\ref{eq_Hdotphonon}) into a procuct of $\gamma_{\bf{q}}$ and $n_{\nu}$ is valid for
strongly confined QDs, where it is justified to treat excitons and biexcitons as product 
states representing uncorrelated electron-hole pairs. However, this factorization does not 
hold in general. For QDs far from the strong confinement limit, e.~g., interface QDs,
measurements \cite{peter:04} indicate strong deviations from Eq.~(\ref{eq_Hdotphonon}).
Details on the phonon coupling can be found in Ref.~\onlinecite{krummheuer:02}.

Finally, the coupling to the light field reads within the common dipole and rotating
wave approximation   
\begin{align}
\label{eq_matrixM}
{H}_{\rm{dot, \, light}} = 
 -\hbar \, [  &F_{\sigma_+}(t) \left( | + \rangle \langle \rm{G} | + | \rm{B} \rangle \langle - | \right) \\ \notag
                     +  &F_{\sigma_-}(t) \left( | - \rangle \langle \rm{G} | + | \rm{B} \rangle \langle + | \right)
                  + \rm{h.c.}] \, ,
\end{align}
where $F_{\sigma_{\pm}} = {\bf{E}}^{(+)}_{\sigma_{\pm}}(t) {\bf{M}} / \hbar$ with 
${\bf{E}}^{(+)}_{\sigma_{\pm}}(t)$ being the positive frequency part of 
the $\sigma_{\pm}$ circularly polarized component of the light field and 
$\bf M$ denoting the ground state exciton transition dipole moment,
that we assume to be identical to the exciton biexciton transition 
dipole moment. Taking the laser frequency to be in resonance with the polaron
shifted exciton transition frequency
$\overline{\Omega} = \Omega-\sum_{\bf{q}} \gamma_{\bf{q}}^2 / \omega_{\bf{q}}$,
we set $F_{\sigma_{\pm}}(t)=f_{\sigma_{\pm}}(t) \exp(-i\overline{\Omega} t)/2$
with real envelope functions $f_{\sigma_{\pm}}(t)$, that we will refer to as
$\sigma_{\pm}$ circularly polarized field strengths. The total field strength
is given by $f(t) = \sqrt{f^2_{\sigma_{+}}(t)+f^2_{\sigma_{-}}(t)}$ and defines
the pulse area $\alpha$ of an applied laser pulse via 
\begin{align}
 \alpha = \int_{-\infty}^{\infty} f(t) \, dt \, . 
\label{alpha}
\end{align}
For linearly polarized laser pulses, the total pulse intensity is equally distributed
between both circularly polarized components, i.~e.,
$f_{\sigma_{+}} \!=\! f_{\sigma_{-}} \!=\! f/\sqrt{2}$, and both spin degenerate
excitons experience the same dynamics. For circularly polarized pulses with
$f \!=\! f_{\sigma_{\pm}}$ and $f_{\sigma_{\mp}} \!=\! 0$, the system reduces for
$V_{\rm{ex}}=0$ to an effective two-level system. All other polarizations we will
refer to as elliptical. Note, that the definition of the pulse area in Eq.~(\ref{alpha})
is such, that for circular polarizations (i.e., a two-level model), a pulse area
of $\alpha = \pi$ inverts the system. \cite{allen:75}

\subsection{Path integral approach}\label{pathint}

To study the combined carrier-phonon dynamics, we use a real-time path integral
approach, that for the four-level system has first been applied in Ref.~\onlinecite{glaessl:12}.
This approach is numerically exact in the sense that there are no approximations
within the model given above, and thus we face no errors beyond well controllable
discretization errors. The reduced electronic density matrix is calculated by taking the
Liouville von Neumann equation for the full density matrix and tracing over the phonon
degrees of freedom, where the time evolution operator is represented as a path integral.
Tracing out the phonon degrees of freedom introduces a memory and leads to a non-Markovian
dynamics for the reduced density matrix, that can be fully accounted for.
A complete description of our algorithm 
can be found in Ref.~\onlinecite{vagov:11}, where we give not only a detailed derivation,
but highlight also specifics of our implementation that are related to the superohmic
character of the carrier-phonon coupling and that are not encountered in implementations
of path integrals for studies with ohmic or subohmic coupling types.
\cite{makri:95a,makri:95b,thorwart:00,thorwart:10}  

The numerically complete treatment within the path integral approach rapidly 
becomes ambitious, when more than two electronic levels are taken into account. 
As explained in Ref.~\onlinecite{vagov:11}, the
number of paths that have to be considered is given by $L^{2(n_c+1)}$, where $L$ is
the number of electronic levels (i.e., here $L=4$) and $n_c$ the number of retarded time
steps that are accounted for. Obviously, keeping $n_c$ fixed, 
the number of paths is squared when one accounts for four instead of two electronic levels, 
and thus the numerical effort increases drastically.
A reduction of the number of paths can be achieved by invoking an
on-the-fly path selection as described in Ref.~\onlinecite{vagov:11} and first 
introduced by Sim.\cite{sim:01} This enables studies with a sufficiently fine time
discretization but at the price of enforcing additional convergence tests, that have
to be performed individually for different parameter ranges.

\subsection{Weak Coupling Theory}\label{perturbative}

In order to better understand the origin of the polarization dependence 
of the phonon-induced damping and some other features to be presented in 
Sec.~\ref{results}, we set up equations of motion for the reduced electronic
density matrix within a weak coupling theory \cite{rossi:02}. A more detailed
description of this perturbative approach that is essentially equivalent 
to a standard Born-Markov master equation approach \cite{breuer:02} is given in the appendix.
Here, we shall only give a short summary.

First, we transform $H$ as given in Eqs.~(\ref{eq_Hfull})-(\ref{eq_matrixM}) from the electronic basis
$\{ |\rm{G}\rangle,|+\rangle,|-\rangle,|\rm{B}\rangle \}$ to the basis spanned by the instantaneous
eigenstates of $H_{\rm{dot}}\!+\!H_{\rm{dot-light}}$, that are usually referred
to as dot-photon dressed states. To keep the formulas as simple as possible,
we concentrate on the special case $\Delta\!=\!V_{\rm{ex}}\!=\!0$. For this choice,
the dressed states are independent on $f_{\sigma_{\pm}}$ as well as on time
and given by ${\bf d_1}=(1,1,1,1)$, ${\bf d_2}=(1,1,-1,-1)$, ${\bf d_3}=(1,-1,-1,1)$
and ${\bf d_4}=(1,-1,1,-1)$ with corresponding eigenenergies $-\hbar \lambda_j$,
where $\lambda_1= f_{+}$, $\lambda_2=f_{-}$, $\lambda_3=- f_{+}$, $\lambda_4=- f_{-}$
and $f_{\pm}=\left(f_{\sigma_+}\pm f_{\sigma_-}\right)/2$.

Setting up equations of motion for the reduced electronic density matrix
$\rho_{ij}=\big\langle |d_i\rangle\langle d_j|\big\rangle$ leads to 
a hierarchy of phonon assisted quantities, that we truncate by factorizing
at the level of two phonon assistances. Next, we integrate the one phonon assisted density
matrix elements performing a Markov approximation and substitute the
resulting expressions back into the equation for the reduced density
matrix. Transforming back to the electronic basis 
$\{ |\rm{G}\rangle,|+\rangle,|-\rangle,|\rm{B}\rangle \}$ yields for example:   
\begin{align}
\label{perturbeqs}
 \partial_t \rho_{12}  
    &= i/2 \, \left[f_{\sigma_+} \left( \rho_{11}\!-\!\rho_{22} \right) +  f_{\sigma_-} \left( \rho_{14}\!-\!\rho_{32} \right) \right] \\ \notag
       & \quad +\pi/4 \, \left[J(f_{\sigma_+}) \left( \rho_{11}\!+\!\rho_{22} \right)
                               -J(f_{\sigma_-}) \left( \rho_{14}\!-\!\rho_{32} \right) \right] \\ \notag
       & \quad -\pi/2 \,\, J(f_{\sigma_+}) \coth \left( \hbar f_{\sigma_+} / 2kT \right) \rho_{12}  \, , 
\end{align}
where we have introduced the phonon spectral density 
\begin{align}
 J(\omega) = \sum_{\bf q} \gamma_{\bf q}^2 \delta(\omega-\omega_{\bf q}) \, .
\end{align}
The main purpose for deriving the above approximate equations in addition
to our numerically complete treatment within the path integral formalism
is to obtain simple closed form expressions for the relevant damping rates 
in terms of the phonon spectral density
 that allow for an explicit discussion of the dependencies on the field
components $f_{\sigma_{\pm}}$. The field strengths enter the damping rates
because the energy conservation condition, that occurs in the Markov limit,
enforces the phonon frequency to match the difference between electronic
dressed state energies $E_{j}=-\hbar\lambda_{j}$, which are directly related
to $f_{\sigma_{\pm}}$. Therefore, the phonon spectral density $J(\omega)$ 
is evaluated at the corresponding transition frequencies. For the special case
$\Delta\!=\!V_{\rm{ex}}\!=\!0$ discussed here, it is important to note
that $J(\omega)$ is only evaluated at $f_{\sigma_+}$ and $f_{\sigma_-}$
[cf.~Eq.~(\ref{perturbeqs}) as an example], i.e., it is the strength of 
the $\sigma_{+}$ and the $\sigma_{-}$ component separately, and not the
total intensity, that matters.

\section{RESULTS}
\label{results}


In the following we will study the electronic dynamics that result from
the interplay between optical excitation and the phonon-induced dephasing.
For our calculations, that have been performed using the path integral
approach, we choose $a_e = 4 \, \rm{nm}$, we set $a_h = 0.87 a_e$, and
assume that initially, the QD is in the unexcited ground state and the
phonons in a thermal state at temperature $T$. $\Delta$ and $V_{\rm{ex}}$
are treated as parameters.
In Sec.~\ref{linear} we concentrate on linearly polarized excitations and
investigate how the laser pulse intensity affects the dynamics. Here, a
special focus is put on comparing the results with those of the widely 
studied two-level model, where biexcitonic excitations are not accounted 
for. Sec.~\ref{elliptical} is devoted to the case of elliptically polarized
excitations and the polarization dependence of the phonon influence.

\subsection{Linearly Polarized Excitations}
\label{linear}

\begin{figure}[ttt]
 \includegraphics[width=8.7cm]{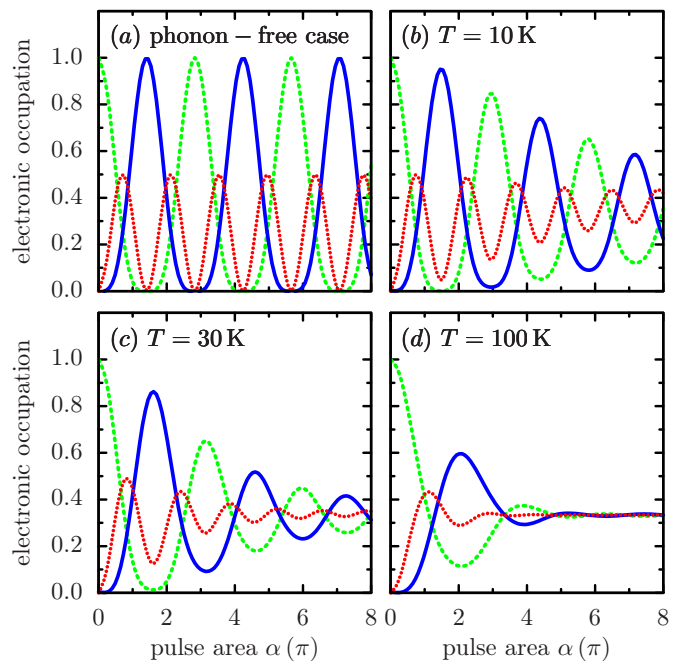}
 \caption{(Color online) Occupation of the ground state $C_{\rm{G}}$ (green dashed line),
          the total single exciton occupation $C_{\rm{E}}$ (red dotted line), and the 
          biexciton occupation $C_{\rm{B}}$ (blue solid line) as a function of the
          applied pulse area $\alpha$ for a linearly polarized rectangular pulse
          of $12 \, \rm{ps}$ duration and $\Delta \!=\! V_{\rm{ex}} \!=\! 0$.
          (a) Displays results for the phonon-free
          case, (b)-(d) show the dynamics in the full model at temperatures of
          (b) $T = 10$, (c) $30$, and (d) $100$~K.
          }
 \label{fig1}
\end{figure}

Let us first concentrate on linearly polarized excitations ($f_{\sigma_+} \!=\! f_{\sigma_-}$)
and consider the Rabi rotation scenario, where the final occupation after the pulse
is recorded as a function of the applied pulse area $\alpha$ as given in
Eq.~(\ref{alpha}) which is most convenient to be studied in experiments. Fig.~\ref{fig1}
shows corresponding results for the ground state occupation $C_{\rm{G}} \!=\! \rho_{11}$
(green dashed line), the total single exciton occupation
$C_{\rm{E}} \!=\! \rho_{22}\!+\!\rho_{33}$ (red dotted line), and the biexciton occupation
$C_{\rm{B}} \!=\! \rho_{44}$ (blue solid line) for a rectangular $12$~ps lasting pulse at
different temperatures for $\Delta \!=\! V_{\rm{ex}} \!=\! 0$.
In the phonon-free case [Fig.~\ref{fig1}(a)], the occupations of
the ground and biexciton state perform undamped oscillations between
$0$ and $1$ with a period of $2\sqrt{2}\pi$, while the total single
exciton occupation does not take values above $0.5$ and oscillates
twice as fast. Recall, that the definition of the pulse area has been
chosen such, that for a two level system the occupation oscillates
with a period of $2 \pi$ and that for linearly polarized excitations
$f_{\sigma_{\pm}}=f/\sqrt{2}$, explaining the additional factor
$\sqrt{2}$ that enters the period of the ground and biexciton state
occupation. Accounting for the carrier-phonon interaction yields a 
damping and a renormalization of the Rabi period. As shown in 
Figs.~\ref{fig1}(b)-\ref{fig1}(d) both effects strongly depend on
temperature. Qualitatively similar results have previously been
reported for electronic two-level systems \cite{foerstner:03,kruegel:05,
vagov:06,glaessl:11}.

\begin{figure}[ttt]
 \includegraphics[width=8.7cm]{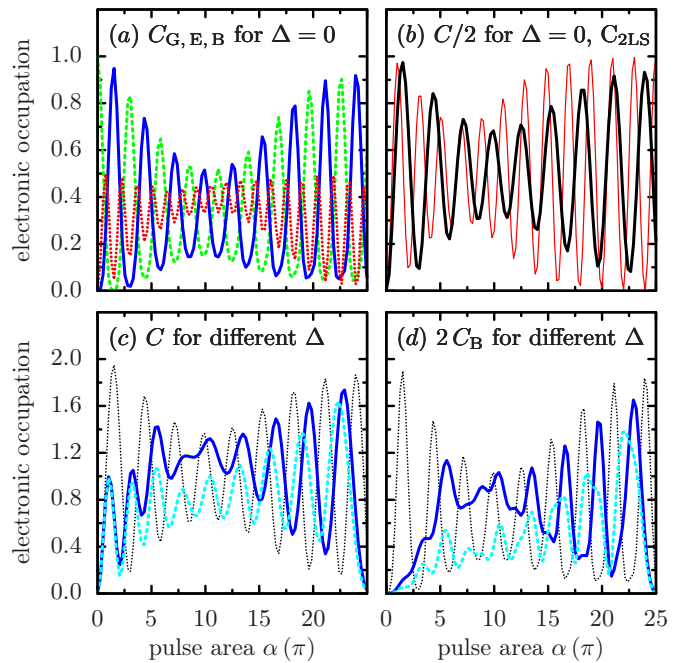}
 \caption{(Color online) Rabi rotation scenarios at $T=10$~K for a linearly
          polarized rectangular pulse of $12$~ps duration.
          (a) $C_{\rm{G}}$ (green dashed line), $C_{\rm{E}}$ (red dotted line),
          and $C_{\rm{B}}$ (blue solid line) for $\Delta=0$~meV.
          (b) Total electronic occupation $C\!=\!C_{\rm{E}}+2C_{\rm{B}}$ (thick black line) and corresponding
          result for a two-level system (red thin line, see text).
          (c) $C\!=\!C_{\rm{E}}+2C_{\rm{B}}$ for $\Delta=0$ (black dotted line), $+1.0$
          (blue solid line), and $-1.0$~meV (blue dashed line).
          (d) Biexciton contribution $2\,C_{\rm{B}}$ to the signals plotted in (c)
          for the same parameters as there.}
 \label{fig2}
\end{figure}

The probably most striking theoretical prediction for the dissipative two-level
dynamics is that the phonon-induced damping should depend nonmonotonically on the
pulse intensity resulting in a reappearance of Rabi rotations at high pulse areas
\cite{vagov:11,glaessl:11}. Recently, first experimental evidence for this
reappearance has been reported \cite{ramsay:10b} for QDs driven
by circularly polarized light, where a two-level model can be applied:
the measured data showed a clear roll-off behaviour at high pulse areas.
Of course, the question arises as to how this scenario appears for linear
polarization, in particular for finite values of the biexciton binding energy,
and which differences can be expected compared to the two-level case.
To answer these questions, different Rabi rotation scenarios are presented
in Fig.~\ref{fig2} for a $12$~ps lasting rectangular pulse at $T=10$~K. 
The rectangular pulse shape, that in experiments can be realized
applying well known pulse-shaping techniques \cite{weiner:95}, has been chosen as
for such pulses the reappearance is more pronounced than for bell-shaped
pulses \cite{glaessl:11}.

Shown in Fig.~\ref{fig2}(a) are Rabi rotations for $\Delta=V_{\rm{ex}}=0$. While
for small pulse areas the damping increases with rising pulse intensity (as also
seen in Fig.~\ref{fig1}), it
decreases at high pulse areas resulting in a reappearance of Rabi rotations. The pulse
area $\alpha_c$ with minimal Rabi rotation amplitude is the same for $C_{\rm{G}}$, $C_{\rm{E}}$ and
$C_{\rm{B}}$ and for the chosen parameters roughly given by $\alpha_c=11\pi$. We stress that
the reappearance of Rabi rotations is different from the collapse and revival phenomenon
within the Jaynes-Cummings model \cite{jaynes:63}, which describes the temporally
periodic carrier dynamics of a two-level system coupled to a single-mode quantized photon field.
For the system considered here, the oscillation amplitudes of all occupations decay
monotonically with time, but with a damping depending nonmonotonically on the applied
pulse area. The physical origin of this nonmonotonic damping, that is reflected in an
undamping of Rabi rotations at high pulse areas, is the resonant nature of the
carrier-phonon interaction \cite{machnikowski:04}: the exciton-phonon coupling is
maximal at intermediate pulse areas, where the frequencies of the electronic
oscillations are resonant with the most strongly coupled lattice modes.

In experimental studies, the final QD occupation after a pulse is often measured
using a photocurrent detection technique, \cite{zrenner:02,ramsay:10,zecherle:10}
where under the action of an applied electric field, the carriers, that have been
created during the exciting pulse, tunnel from the QD resulting in a photocurrent.
For an exciton-biexciton system, this photocurrent is proportional to the weighted
sum $C:=C_{\rm{E}} + 2C_{\rm{B}}$ of the generated single exciton and biexciton populations. In
the following, we will concentrate on this total electronic occupation instead of
the single contributions $C_{\rm{E}}$ or $C_{\rm{B}}$ that in experiments are much harder to
extract individually.

Plotted in Fig.~\ref{fig2}(b) is $C/2$ for $\Delta\!=\!V_{\rm{ex}}\!=\!0$ (thick black line)
together with the exciton occupation $C_{\rm{2LS}}$ of a two-level system (thin red line),
that for $V_{\rm{ex}} = 0$ we can model by choosing circularly polarized laser light. 
For the four-level case,
the maximal signal is not only twice as high as in the two-level case (mind the scaling
factor of 0.5 for $C$) but the pulse area with minimal amplitude, marking the beginning
of the reappearance, is also considerably higher. This difference can be understood
with the help of Eq.~(\ref{perturbeqs}): the phonon-induced damping is expressed in
terms of the phonon spectral density $J(\omega)$ that rises for small frequencies and
peaks at a certain frequency $\overline{\omega}$ before it eventually decreases
\cite{vagov:11,hodgson:08}. The important point to note is that for the chosen
parameters $J$ is not evaluated at the total field strength $f$ but at the circularly
polarized field strengths $f_{\sigma_+}$ and $f_{\sigma_-}$ (which are identical for
linear polarization). Further, for a given pulse area $\alpha$, the circularly 
polarized field strengths of a linearly polarized pulse are by a factor of $\sqrt{2}$
smaller than the field strength of a circularly polarized pulse.
Thus, to reach the field strength, where $J$ and thereby the phonon-induced damping is
maximal, higher pulse areas are needed for linearly polarized pulses than for circularly
polarized pulses and hence, the reappearance starts later in the linear case.
For the same reason, higher pulse areas are needed to
fully restore the maximal Rabi rotation amplitude.

Next, in Fig.~\ref{fig2}(c), we shall discuss the influence of the biexciton binding
energy, which, depending on details of the quantum dot geometry, may take negative as
well as positive values. While for $\Delta = 0$ (black dotted line) the dependence of
$C$ on $\alpha$ is similar to that of the two-level case, Rabi scenarios
become much more complex for finite biexciton binding energies, where they qualitatively
differ from the two-level results. For $\Delta=-1.0$
(blue solid line) or $+1.0$~meV (blue dashed line) there is still a reappearance
in the sense that first the Rabi rotation amplitude decreases with increasing pulse
area before it eventually rises again, but in contrast to $\Delta=0$ the signal
consists of a superposition of different transition frequencies and the mean value varies
with $\alpha$. For small pulse areas, the mean value is less than one and the signal does almost
not depend on the sign of $\Delta$. For intermediate pulse areas, it
increases and we observe a pronounced difference between $\Delta =-1.0$ and $+1.0$~meV,
whereas for high pulse areas, both binding energies result again in similar
oscillations with a mean value of approximately one. Further, there is a clearly visible
increase of the oscillation period with rising pulse areas. 

To understand these features, it is helpful to consider the biexciton contribution
$2C_{\rm{B}}$ to the total photocurrent as shown in Fig.~\ref{fig2}(d). For small pulse areas,
the driving field strengths $f_{\sigma_{\pm}}$ are much smaller than $|\Delta/\hbar|$.
From analytic expressions for driven two-level systems without a coupling to phonons
\cite{allen:75}, it is well known that off-resonant transitions become unlikely, when
the frequency detuning is larger than the driving field strength and
that the dynamics does not depend on the sign of the detuning. Accounting for
phonons, these statements remain approximatively valid as long as the time-scales
considered are much shorter than the time-scale, on which the system is driven
to a stationary state \cite{glaessl:11b}, which is fulfilled for small pulse areas. 
Thus, the biexciton is hardly excited [cf.~Fig.~\ref{fig2}(d)] in the regime of low
pulse intensities and $C$ basically given by $C_{\rm{E}}$. With increasing pulse area, the
generation of the biexciton becomes more and more likely and $C_{\rm{B}}$ starts to contribute,
which is reflected in an increase of the mean value of $C$. At intermediate pulse
areas, the contribution of $C_{\rm{B}}$ is noticeable but depends strongly on the sign of
$\Delta$. The latter can be explained as follows: as already mentioned above, the
carrier-phonon interaction is maximal at intermediate pulse areas, where
the system is quite rapidly driven to a stationary non-equilibrium state. As shown
in Ref.~\onlinecite{glaessl:12}, the stationary occupations depend
on the question whether dark superposition states are realized or not and are also strongly
dependent on $\Delta$ and $T$. For the given temperature of $T=10$~K and a driving field strength
of $f=2.0 \, \rm{ps}^{-1}$ [corresponding to pulse areas for which the Rabi rotation
amplitude is minimal in Fig.~\ref{fig2}(c)], the stationary biexciton
occupation is roughly $0.55$ for $\Delta=1$~meV, while for $\Delta=-1.0$~meV it is
strikingly less and roughly given by $0.14$. Although we never reach a stationary
state within the considered pulse duration of $12$~ps, fingerprints of the relaxation
process become visible in the regime of an enhanced carrier-phonon interaction as
given at intermediate pulse areas and lead to the strong dependence on the sign of
$\Delta$. At high pulse areas, the driving field strengths are eventually significantly
larger than $|\Delta/\hbar|$, the biexciton can be efficiently excited and due to the
decreasing phonon influence with increasing pulse area the system is more and more
far from becoming stationary, which results in a diminishing dependence on the sign
of $\Delta$. Finally, the increasing period of the oscillations of $C$ with rising
$\alpha$ can be explained by the increasing contribution of $C_{\rm{B}}$ that shows a
considerably longer period than $C_{\rm{E}}$ as already discussed in Fig.~\ref{fig1} and
also shown in Fig.~\ref{fig2}(a).


\subsection{Elliptically Polarized Excitations}
\label{elliptical}

So far, we have studied the exciton-biexciton dynamics under linearly polarized pulses.
It was shown that the coupling between carriers and phonons strongly depends on the
pulse intensity and that Rabi rotation scenarios for finite $\Delta$ are qualitatively
different from those of two-level models, while they are simliar for $\Delta \sim 0$. 
In what follows, we shall demonstrate a remarkable polarization dependence of the
phonon-induced damping in an exciton-biexciton system, which has obviously no analogue
in a two-level system.

\begin{figure}[ttt]
\includegraphics[width=8.7cm]{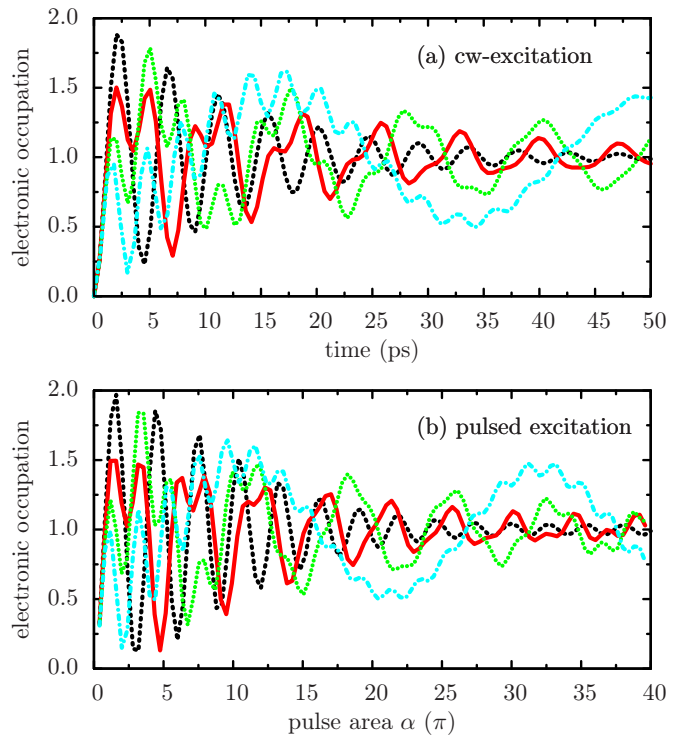}
 \caption{(Color online) Totally created charge $C\!=\!C_{\rm{E}}+2C_{\rm{B}}$ 
          for different polarization parameters $\gamma$ (see text):
          $\gamma = 1.0$ (black dashed line), $0.5$ (red solid line),
          $0.3$ (green dotted line), and $0.1$ (blue dashed-dotted line)
          at $T=10$~K and for $V_{\rm{ex}}=\Delta=0$.
          (a) shows the dynamics under cw-excitation for a total 
          field strength of $f=2.0 \, \rm{ps}^{-1}$ as a function of time,
          (b) displays the final QD occupation after a Gaussian shaped pulse 
          with a FWHM of $30\,\rm{ps}$ as a function of the applied pulse area.}
 \label{fig3}
\end{figure}

To quantify the ellipticity of the polarization, we first introduce a polarization
parameter $\gamma$ via $f_{\sigma_-}\! = \!\gamma f_{\sigma_+}$. For $\gamma\!=\!1$,
we obtain linearly polarized light, while a circularly polarized excitation is 
realized for $\gamma \!=\! 0$.  
Plotted in Fig.~\ref{fig3}(a) is the total electronic occupation $C\!=\!C_{\rm{E}}\!+\!2C_{\rm{B}}$
as a function of time under cw-excitation for polarizations with $\gamma\!=\!1$ (black dashed line),
$0.5$ (red solid line), $0.3$ (green dotted line), and $0.1$
(blue dotted-dashed line), respectively. The total field strength has been fixed
to $f=2.0 \, \rm{ps}^{-1}$, i.e., the pulse intensity is the same for all
displayed results, and we have chosen a vanishing
biexciton binding energy $\Delta$ as well as a vanishing exchange interaction
$V_{\rm{ex}}$, as for $\Delta=V_{\rm{ex}}=0$ an interpretation based on the simple
rates derived in Sec.~\ref{perturbative} for this case can be made.
For all excitation conditions, the total electronic occupation
performs damped Rabi oscillations. But remarkably, the phonon-induced damping
strongly depends on the polarization of the excitation. While the damping is
strongest for linearly polarized light, the signal is the less damped the 
closer the polarization is to be circular. For $\gamma=0.1$, corresponding
to an almost circularly polarized excitation, the oscillation amplitude is
after $50$~ps more than 15 times larger than for linear polarization
($\gamma\!=\!1$), where the oscillation amplitude is after $50$~ps strongly
reduced and the system already close to becoming stationary. 
As the carrier-phonon coupling as given by $H_{\rm{dot-ph}}$ is identical for 
$\sigma_+$ and $\sigma_-$ excitons, a polarization dependent damping of Rabi
rotations is at first a very surprising finding. In light of a spin-independent
exciton-phonon interaction one could rather expect that the damping is 
fully determined by the laser intensity and should thus depend only on $f$
or other quantities related to the total intensity like
$f_{\sigma_{+}}+f_{\sigma_{-}}$. However, our results clearly demonstrate
that this is not the case and reveal a strong polarization dependence of 
the phonon-induced damping.

For the discussion of this remarkable finding, cw-excitations
are most advantageous, as for this choice the field strengths $f_{\sigma_{\pm}}$
are time-independent and an analysis based on field-strength dependent damping
rates as given below is particularly simple. However, in experiments it is a
challenge to obtain time-resolved traces of the electronic occupation as presented
in Fig.~\ref{fig3}(a) and in most experimental studies 
pulsed excitations are used, where the final occupation is recorded as a function
of the applied pulse area. We would like to stress, that the polarization dependence
of the phonon-induced damping is also clearly seen for the latter scenario. This is
exemplarily shown in Fig.~\ref{fig3}(b), where the final occupation is plotted
as a function of the pulse area for a Gaussian shaped pulse with a pulse duration
of $30\,\rm{ps}$ (FWHM):
the reduction of the Rabi rotation amplitude with rising pulse area is 
the less the closer the polarization is to the circular limit, reflecting a
decreased damping in the time domain.

The most convenient way to think of phonon-mediated relaxation is that
the laser couples the excitonic states to form optically dressed states,
that represent stable quasiparticles \cite{glaessl:11b}. 
As discussed in Sec.~\ref{perturbative}, the damping of Rabi oscillations
is determined by the phonon spectral density evaluated at the transition
frequencies between the dressed states, which for $\Delta\!=\!V_{\rm{ex}}\!=\!0$ 
are given by $f_{\sigma_{+}}$, $f_{\sigma_{-}}$, 
$f_{\sigma_{+}}\!-\!f_{\sigma_{-}}$ and $f_{\sigma_{+}}\!+\!f_{\sigma_{-}}$,
but it turns out that terms involving $f_{\sigma_{+}}\!+\!f_{\sigma_{-}}$
or $f_{\sigma_{+}}\!-\!f_{\sigma_{-}}$ drop out.
This result explains the strong dependence of the damping on the
polarization seen in Fig.~\ref{fig3} as follows (for simplicity,
we assume again cw-excitation): while for linear
polarization both circularly polarized components are given by $f/\sqrt{2}$,
for elliptical polarizations $f_{\sigma_{+}}$ is larger and $f_{\sigma_{-}}$
smaller than this value (without loss of generality we can assume
$f_{\sigma_{+}}$ to exceed $f_{\sigma_{-}}$, corresponding to $\gamma < 1$).
Depending on the total laser intensity, we can then encounter two situations:
either $J(f_{\sigma_{-}}) < J(f_{\sigma_{+}})$ and the $\sigma_-$ driven
subsystem experiences a weaker damping than in the case of an equally strong
linear driving [which is fulfilled for the driving strength of 
$f=2.0 \, \rm{ps}^{-1}$ in Fig.~\ref{fig3}(a)], or
$J(f_{\sigma_{+}}) < J(f_{\sigma_{-}})$ and the damping of the $\sigma_+$
driven subsystem is less strong. In any case, one subsystem experiences a
weaker damping than in the linear case, which results in a weaker damping
of the total electronic occupation, to which both subsystems contribute. 
From this reasoning it follows straightforwardly that for elliptically
polarized excitations the difference of the damping compared to the linear
case should monotonically increase with rising ellipticity, which is 
indeed confirmed by the results shown in Fig.~\ref{fig3}.

For elliptical polarizations close to the circular limit, the carrier dynamics in one
subsystem experiences eventually hardly any damping and $C$ oscillates with large
amplitudes over a long time-interval. In Fig.~\ref{fig3}(a) this is nicely illustrated for
$\gamma=0.1$ (blue dashed-dotted line).
First, the signal rapidly oscillates. However, after roughly 20~ps these fast
oscillations have decayed. Subsequently, the dynamics changes qualitatively
and is dominated by oscillations with a much longer period that are only 
weakly damped. While the fast oscillations 
reflect the dynamics in the $\sigma_+$ driven subsystem, the much less damped
oscillations with a longer period stem from the $\sigma_-$ driven subsystem.
As for the chosen polarization $f_{\sigma_-}$ is roughly ten times smaller than
$f_{\sigma_+}$, the oscillations in the $\sigma_-$ driven subsystem reveal
a roughly ten times longer period and experience a considerably less 
damping, because here $J(f_{\sigma_{-}}) \ll J(f_{\sigma_{+}})$.
This characteristic change in the dynamics is also expressed
in corresponding results for pulsed excitations, cf. Fig.~\ref{fig3}(b).

The limit of a strictly circular polarization with $\gamma = 0$ is special:
for this excitation condition, one subsystem is driven with the total intensity,
while the other subsystem is not driven at all. In consequence, while
for almost circularly polarized light the damping seen in the oscillations
of the total occupation decreases with rising ellipticity and the time to
become stationary rises without any limit for $\gamma \to 0$, this time is
finite for circularly polarized light as discussed in more detail in Figs.~2(b) and (c) of
Ref.~\onlinecite{glaessl:12}.

\begin{figure}[ttt]
\includegraphics[width=8.7cm]{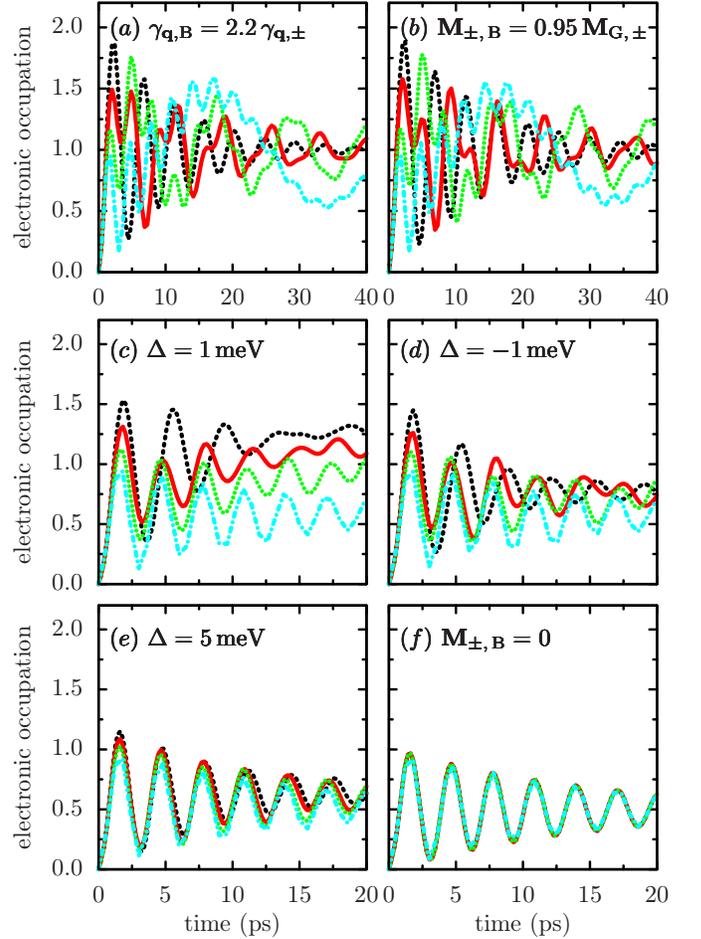}
 \caption{(Color online) Totally created charge $C\!=\!C_{\rm{E}}\!+\!2C_{\rm{B}}$ 
          under cw-excitation as a function of time at $T=10$~K,
          $f=2.0 \, \rm{ps}^{-1}$ and $V_{\rm{ex}}=0$ for different polarizations 
          with $\gamma = 1.0$ (black dashed line), $0.5$ (red solid line),
          $0.3$ (green dotted line), and $0.1$ (blue dashed-dotted line)
          for situations, in which the relevant damping rates do not separatively
          depend on $f_{\sigma_+}$ or $f_{\sigma_-}$ (see text).}
 \label{fig4}
\end{figure}

It should be noted, that the finding that the damping rate depends separately
on $f_{\sigma_{+}}$ and $f_{\sigma_{-}}$ strictly holds only in the strong confinement
limit as defined by Eqs.~(\ref{eq_Hfull})-(\ref{eq_matrixM}) and in the limiting
case $\Delta\!=\!V_{\rm{ex}}\!=\!0$ discussed so far. In the phonon-free case,
it is easily seen that for $\Delta\!=\!V_{\rm{ex}}\!=\!0$ the model defined
by Eqs.~(\ref{eq_Hfull})-(\ref{eq_matrixM}) can be decomposed into a product
of two two-level systems, each evolving on its Bloch sphere with a Rabi
frequency given by $f_{\sigma_{+}}$ or $f_{\sigma_{-}}$, respectively. Thus,
in this case we have two subsystems that evolve independently and react
separately either on the values of $f_{\sigma_{+}}$ or $f_{\sigma_{-}}$.
In a four-level system composed of two independent two-level systems each
coupled to phonons, the phonon coupling of the two-exciton state must
necessarily be twice the exciton-phonon coupling. Although this necessary
condition is fulfilled in the exciton-biexciton system studied so far, we
are actually not dealing with strictly independent dynamics on two
Bloch-spheres as the two spheres are coupled to the same phonon modes.
The resulting interdependence of the dynamics on the two Bloch spheres
can be seen, e.g., from the fact that for linearly polarized driving the
stationary biexciton occupation reached at long times is not given by $1/4$ 
(which would be the product of the stationary values of $1/2$ reached in any
of the $\sigma_{\pm}$ coupled two-level systems) but 1/3. \cite{glaessl:12}
Nevertheless, even though a strict factorization into two independently
evolving subsystems does not hold, a phonon-coupling of the biexciton state
$\gamma_{{\bf{q}},\rm{B}}$ of exactly twice the exciton phonon-coupling 
$\gamma_{{\bf{q}},\pm}$ realizes a distinguished situation as in this case the
phonon induced damping rates depend on $f_{\sigma_{+}}$ and $f_{\sigma_{-}}$
separately. However, this property is lost, when either 
$\gamma_{{\bf{q}},\rm{B}}\neq 2 \gamma_{{\bf{q}},\pm} $ or when the biexciton
binding energy or the exchange interaction take finite values, or when the
dipole moments associated with the ground state exciton transition
${\bf{M}}_{\rm{G},\pm}$ and the exciton biexciton transition
${\bf{M}}_{\pm,\rm{B}}$ are different. In each of these cases terms involving
$f_{\sigma_+}+f_{\sigma_-}$ or $f_{\sigma_+}-f_{\sigma_-}$ enter the equations
derived within the weak coupling theory. It turns out, however, that for many 
typical situations with deviations from the limiting case discussed so far, the
amplitudes of the additional terms are small and the polarization dependence
is still close to the results shown in Fig.~\ref{fig3}. This is illustrated
in Fig.~\ref{fig4} and shall be discussed in the remainder of the paper.

In real systems, ${\bf{M}}_{\pm,\rm{B}}$ differs from ${\bf{M}}_{\rm{G},\pm}$
due to deviations from the strict strong confinement limit, where the
multi-pair wave functions factorize in single particle wave functions for
electrons and holes. Typical values observed in experiments indicate, that
${\bf{M}}_{\pm,\rm{B}}$ is roughly $5\%$ smaller than ${\bf{M}}_{\rm{G},\pm}$.
\cite{stufler:06, boyle:08} For the same reason,
$\gamma_{{\bf{q}},\rm{B}} \neq 2\gamma_{{\bf{q}},\pm}$
is possible. Figs.~\ref{fig4}(a) and \ref{fig4}(b) show results for calculations
assuming ${\bf{M}}_{\pm,\rm{B}}=0.95\,{\bf{M}}_{\rm{G},\pm}$ and
$\gamma_{{\bf{q}},\rm{B}}=2.2\,\gamma_{{\bf{q}},\pm}$, respectively, for the same
polarizations as in Fig.~\ref{fig3}. It is clearly seen that the polarization dependence
is only weakly influenced by these deviations from the strong confinement limit.

We have also performed calculations (not shown) accounting for a finite exchange
interaction of $V_{\rm{ex}}=30 \, \mu$eV, which represents a typical value.
Although $V_{\rm{ex}}$ yields a direct coupling of the two subspaces driven
by $\sigma_{\pm}$ polarized light, we found only very small changes compared
to $V_{\rm{ex}}=0$, except for strict circular polarization \cite{glaessl:12} 
or in the regime of nearly circularly polarized excitations,
when the time-scale on which the total system is driven to a stationary 
state increases drastically and exceeds $h/V_{\rm{ex}}$. 
 
Finally, let us turn to the influence of finite values of the biexciton 
binding energy, as illustrated in Figs.~\ref{fig4}(c) and \ref{fig4}(d) for 
$\Delta = 1.0$ and $-1.0$~meV, respectively. At first one may notice that
the mean value of the oscillations decreases with rising ellipticity of 
the polarization. This lowering of the total signal is due to the fact, that
the biexciton generation becomes more unlikely, when $f_{\sigma_-}$ decreases,
as in this case the exciton biexciton transition given
an exciton generated by $\sigma_+$ polarized light becomes less probable.
In addition, there is no longer a cross-over between fast and slow
oscillations for almost circularly polarized excitations. 
However, despite these differences compared to the case of a
vanishing biexciton binding energy, we find a very similar behavior
as discussed so far: the decay of the oscillation amplitude is the
less pronounced the more far the 
excitation is from the linear case, clearly illustrating a strong
dependence of the phonon-induced damping on the polarization of the
driving laser light. This picture changes, however, for larger $\Delta$,
which are still feasible, as shown in Fig.~\ref{fig4}(e) for 
$\Delta = 5$ meV. Here, the polarization dependence is drastically 
suppressed. As the coupling to the biexciton decreases with rising
$\Delta$, this indicates, that an efficient coupling to the biexciton
is of fundamental importance for the polarization dependence
of the damping. Indeed, when we completely decouple the biexciton 
from the dynamics by setting by hand ${\bf{M}}_{\pm,\rm{B}}=0$, then the
polarization dependence vanishes altogether and the phonon-induced
damping is fully determined by the pulse intensity [cf. Fig.~\ref{fig4}(f),
where all four lines coincide]. Only when the biexciton is coupled,
there is the possibility to essentially drive two different Bloch
spheres with individual Rabi frequencies within the four-level system
considered here, giving rise to the polarization dependence of the 
phonon-induced damping as explained above.

\section{CONCLUSIONS}\label{conclusions} 

In summary, we have presented a numerically complete analysis of the
phonon impact on optically driven exciton-biexciton QD systems. 
Rabi rotation scenarios for finite biexciton binding energies were shown
to differ significantly from the two-level case and offer
manifold insights in the dynamics of an exciton-biexciton system, 
including the interplay between off-resonant driving and quantum
dissipative relaxation. In addition, we have demonstrated that
although the carrier-phonon interaction is independent of the spin,
the phonon-induced damping does not only depend on the pulse
intensity but is for typical parameters also strongly affected
by the polarization of the driving laser pulse. While the damping
is strongest for linearly polarized excitations, it is the more
reduced the closer the polarization is to the circular limit. 
For almost circularly polarized excitations and small biexciton
binding energies, the Rabi scenario changes strongly in the
course of time reflecting different dynamics in the $\sigma_{\pm}$ 
driven subsystems. The origin of the polarization dependent
phonon-induced damping is the possibility to drive 
two Bloch spheres 
essentially each with its own Rabi-frequency $f_{\sigma_{\pm}}$
that are determined
by the polarization, together with the fact that it is the
Rabi frequency that determines the damping in a two-level system.
This possibility is intimately related to an efficient coupling
to the biexciton state, which highlights the pivotal role of
this two-pair state for the discussed dynamics.

\section{ACKNOWLEDGMENTS}

M.~G. gratefully acknowledges financial support by the Studienstiftung
des Deutschen Volkes. We also highly appreciate many fruitful 
discussions with T.~Kuhn.

\section{Appendix}

In this appendix, we give a short description of the weak coupling theory
presented in Sec. \ref{perturbative}  concentrating on the limit of strong
electronic confinement as defined by Eqs.~(\ref{eq_Hfull})-(\ref{eq_matrixM})
and the special case $\Delta\!=\!V_{\rm{ex}}\!=\!0$. 
The Heisenberg equations of motion for the electronic operators 
$\hat{\overline{\rho}}_{ij} = |d_{i}\rangle\langle d_{j}|$
read
\begin{align}
\label{p1}
 \partial_t \hat{\overline{\rho}}_{ij} = \frac{i}{\hbar} [{H},\hat{\overline{\rho}}_{ij}] \, ,
\end{align}
where the overbar indicates, that we are working in the dressed state basis.
By taking expectation values we obtain equations of motion for the elements of the
reduced density matrix $\overline{\rho}_{ij}=\langle \hat{\overline{\rho}}_{ij}\rangle$,
where single phonon assisted density matrix elements of the form
$\langle \hat{\overline{\rho}}_{i,j} b^{\dag}_{\bf q} \rangle$ or
$\langle \hat{\overline{\rho}}_{i,j} b_{\bf q}        \rangle$ enter.
Setting up equations of motion for these single phonon assisted quantities, in turn,
introduces double phonon assistances. In order to truncate the resulting hierarchy
of higher order phonon assisted quantities, we set 
\begin{subequations}
\begin{align}
 \langle\hat{\overline{\rho}}_{ij}b_{\bf q}^{\dag} b_{\bf q'}\rangle &= \overline{\rho}_{ij} \, \langle b_{\bf q}^{\dag} b_{\bf q'}\rangle 
                                           = \overline{\rho}_{ij} \, n_{\bf q} \, \delta_{\bf q,q'} \\
 \langle\hat{\overline{\rho}}_{ij}b_{\bf q} b_{\bf q'}^{\dag}\rangle &= \overline{\rho}_{ij} \, \langle b_{\bf q} b_{\bf q'}^{\dag}\rangle 
                                           = \overline{\rho}_{ij} \, (n_{\bf q}+1) \, \delta_{\bf q,q'} \\
 \langle\hat{\overline{\rho}}_{ij}b_{\bf q} b_{\bf q'}\rangle        &= \langle\hat{\overline{\rho}}_{ij}b_{\bf q}^{\dag} b_{\bf q'}^{\dag}\rangle
                                           = 0 \, ,
\end{align}
\end{subequations}
where $n_{\bf q}\! =\! 1/(\exp(\hbar \omega_{\bf q} / k_B T)\!-\!1)$
denotes the Bose-Einstein occupancy at temperature $T$.
By doing so, one obtains for the single phonon
assisted elements a set of closed equations. Exemplarily, we have
\begin{align}
\label{example}
 \partial_t \langle\hat{\overline{\rho}}_{12}b_{\bf{q}}\rangle =
            &i (-\omega_{\bf q}+\lambda_{21}) \langle\hat{\overline{\rho}}_{12}b_q\rangle \\ \notag
            &+ \frac{i\gamma_{\bf q}}{2} [-2\overline{\rho}_{12} -n_{\bf q}    (\overline{\rho}_{22}+\overline{\rho}_{42}) \\ \notag
            & \qquad \quad  +(n_{\bf q}+1)(\overline{\rho}_{11}+\overline{\rho}_{13})] \, . 
\end{align}
Integrating these equations yields integrals of the form
\begin{align}
 \int_{0}^{\infty}  \!\!\!\! \exp(i\omega \tau)  \overline{\rho}_{ij}  (t-\tau) d\tau \, .
\end{align}
Next, we perform a Markov approximation by identifying slowly varying variables 
and taking them out of these memory integrals. While occupations can be 
regarded as slowly varying, this is usually not the case for coherences, that are
generally rapidly changing quantities. However, as their free oscillation frequency
is given by $\lambda_{ji} := \lambda_j-\lambda_i$, we can set
$\overline{\rho}_{ij}(t) = \exp(i\lambda_{ji}t) \, \tilde{\overline{\rho}}_{ij}(t)$,
where $\tilde{\overline{\rho}}_{ij}(t)$ can be expected to vary slowly. Taking these
variables out of the integral yields
\begin{align}
 \int_{0}^{\infty} \!\!\!\!\! \exp(i\omega \tau)  \overline{\rho}_{ij}  (t\!-\!\tau) d\tau
  = \overline{\rho}_{ij}(t) \!\!  \int_{0}^{\infty}  \!\!\!\!\! \exp[i(\omega\!-\!\lambda_{ji})\tau] d\tau
\end{align}
and the remaining integrals can be easily evaluated using
\begin{align}
\int_{0}^{\infty} \!\!\!\! \exp(i\omega \tau) d\tau = \pi \delta(\omega) + i \frac{P}{\omega},
\end{align}
where $P$ denotes Cauchy's principal value. It is well known that the principal value terms
only lead to a small renormalization of the system Hamiltonian but do not directly contribute
to dephasing.\cite{breuer:02} Therefore, we neglect these terms and obtain: 
\begin{align}
\label{example2}
 \langle\hat{\overline{\rho}}_{12}b_{\bf{q}}\rangle = \frac{i \pi \gamma_{\bf q}}{2}
              [ &-2\overline{\rho}_{12}\delta(\omega_{\bf{q}}) 
                 - n_{\bf{q}} \overline{\rho}_{22}\delta(\omega_{\bf{q}}+\lambda_{12}) \\ \notag
                &- n_{\bf{q}} \overline{\rho}_{42}\delta(\omega_{\bf{q}}+\lambda_{14}) \\ \notag
                &+ (n_{\bf{q}}+1) \overline{\rho}_{11}\delta(\omega_{\bf{q}}+\lambda_{12}) \\ \notag
                &+ (n_{\bf{q}}+1) \overline{\rho}_{13}\delta(\omega_{\bf{q}}-\lambda_{23})      ] \, . 
\end{align}

Here, only $\lambda_{14}=\lambda_{23}=f_{\sigma_+}$ and 
$\lambda_{12}=\lambda_{43}=f_{\sigma_-}$ appear, although in Eq.~(\ref{example})
also combinations like $\lambda_{24} = f_{\sigma_+}-f_{\sigma_-}$ (via the free
oscillation of $\langle\overline{\rho}_{42}\rangle$) entered, because the
corresponding contributions cancel out. 
It is an important point to note, that this holds for all other phonon-assisted
density matrix elements as well: from here on only $\delta (\omega-f_{\sigma_+})$
and $\delta (\omega-f_{\sigma_-})$ enter the resulting equations. Terms involving
$\delta [\omega-(f_{\sigma_+}+f_{\sigma_-})]$ or
$\delta [\omega-(f_{\sigma_+}-f_{\sigma_-})]$ are absent due to the structure of
$H_{\rm{dot-light}}$ and $H_{\rm{dot-phonon}}$ in the special case considered here.
Introducing the phonon spectral density  
\begin{align}
 J(\omega) = \sum_{\bf q} \gamma_{\bf q}^2 \delta(\omega-\omega_{\bf q})
\end{align}
and substituting the expressions for the single phonon assisted density matrix
elements into Eq.~(\ref{p1}) eventually leads to a closed set of equations of
motion for the elements of the reduced electronic density matrix.

We mention, that the weak coupling theory derived here yields the same result as a four-level
extension of the master-equation approach that has been presented in Ref.~\onlinecite{ramsay:10}
for the case of an electronic two-level system. For two-level systems the latter approach has been compared to a much
more advanced variational master-equation approach in Ref.~\onlinecite{mccutcheon:11}, which in turn
has been shown to agree well with numerically exact path integral calculations for not too high
temperatures and standard GaAs parameters, that represent a weak carrier-phonon coupling regime.
We checked that in the four-level case, the deviations of the weak coupling theory to numerically
exact calculations are qualitatively similar as in the two-level case and refer to Ref.~\onlinecite{mccutcheon:11}
for a comparison. Readers interested in the validity ranges of approximate methods in the regimes
of strong carrier-phonon couplings or high temperatures are
referred to Refs.~\onlinecite{glaessl:11b} and~\onlinecite{vagov:11b}, where path integral 
calculations have been compared to a second-order and fourth-order correlation expansion, 
clearly demonstrating the superiority of a numerically exact approach, accounting fully for all non-Markovian
effects and arbitrary multiphonon processes.


\end{document}